\documentclass[12pt]{article}
\usepackage[margin=1in]{geometry}                
\usepackage{graphicx}
\usepackage{amssymb}
\usepackage{amsmath,bm}
\usepackage{epstopdf}
\usepackage{graphicx}
\usepackage{ulem}
\usepackage{color}
\usepackage{hyperref}
\usepackage{subfigure,enumerate}
\usepackage{booktabs}

\usepackage{natbib}

\def\spacingset#1{\renewcommand{\baselinestretch}%
{#1}\small\normalsize} \spacingset{1}
\title{}
\author{}

\def\pp{\mbox{Poisson}}

\def\gp{\mathcal{GP}}

\newcommand{\cm}[1]{\ignorespaces}

\def\bfa{\mathbf a}
\def\bfb{\mathbf b}

\def\bfx{\mathbf x}

\def\bfz{\mathbf z}
\def\bfs{\mathbf s}

\def\bfS{\mathbf S}

\def\bfI{\mathbf I}
\def\bfX{\mathbf X}

\def\bfbeta{\boldsymbol \beta}
\def\bftheta{\boldsymbol \theta}

\def\bfmu{\boldsymbol\mu}
\def\bfSigma{\boldsymbol\Sigma}

\def\bfzeta{\boldsymbol\zeta}
\def\bfphi{\boldsymbol\phi}
\def\bfzero{\boldsymbol 0}

\def\cA{\mathcal A}

\def\cT{\mathcal T}

\def\cR{\mathcal{R}}
\def\cD{\mathcal D}

\def\mbR{\mathbb R}

\def\md{\mathrm d}

\def\mN{\mathrm{N}}

\def\mU{\mathrm{U}}
\def\mIG{\mathrm{IG}}

\def\diag{\mathrm{diag}}

\def\cA{\mathcal{A}}

\def\rT{\mathrm T}

\newcommand{\blind}{1}

\def\boxit#1{\vbox{\hrule\hbox{\vrule\kern6pt\vbox{\kern6pt#1\kern6pt}\kern6pt\vrule}\hrule}}

\begin{document}

\if1\blind
{
  \title{\bf What Influences the Field Goal Attempts of Professional Players? Analysis of Basketball Shot Charts via Log Gaussian Cox Processes with Spatially Varying Coefficients}
  \author{Jiahao Cao, Qingpo Cai, Lance A. Waller,  DeMarc A. Hickson,\\ Guanyu Hu, and Jian Kang \thanks{Jiahao Cao and Qingpo Cai are joint first authors. Guanyu Hu and Jian Kang are  joint corresponding authors}
     }
  \maketitle
} \fi

\if0\blind
{
  \bigskip
  \bigskip
  \bigskip
  \begin{center}
    {\LARGE\bf What Influences the Field Goal Attempts of Professional Players? Analysis of Basketball Shot Charts via Log Gaussian Cox Processes with Spatially Varying Coefficients}
\end{center}
  \medskip
} \fi

\bibliographystyle{asa}
\bibpunct{(}{)}{,}{a}{}{;}

\begin{abstract}
Basketball shot charts provide valuable information regarding local patterns of in-game performance to coaches, players, sports analysts, and statisticians. The spatial patterns of where shots were attempted and whether the shots were successful suggest options for offensive and defensive strategies as well as historical summaries of performance against particular teams and players. The data represent a marked spatio-temporal point process with locations representing locations of attempted shots and an associated mark representing the shot's outcome (made/missed). Here, we develop a Bayesian log Gaussian Cox process model allowing joint analysis of the spatial pattern of locations and outcomes of shots across multiple games. We build a hierarchical model for the log intensity function using Gaussian processes, and allow spatially varying effects for various game-specific covariates. We aim to model the spatial relative risk under different covariate values. For inference via posterior simulation, we design a Markov chain Monte Carlo (MCMC) algorithm based on a kernel convolution approach. We illustrate the proposed method using extensive simulation studies. A case study analyzing the shot data of NBA legends Stephen Curry, LeBron James, and Michael Jordan highlights the effectiveness of our approach in real-world scenarios and provides practical insights into optimizing shooting strategies by examining how different playing conditions, game locations, and opposing team strengths impact shooting efficiency.
\noindent%

\end{abstract}
\newpage


\spacingset{2}
\section{Introduction}
In recent years, there has been a growing interest in the quantitative analysis of sports data, reflecting the expanding role of analytics in shaping decisions within the sports industry. For instance, national polls rank teams in college football and influence selections for post-season bowl games and championship playoffs. The abundance of data related to various aspects of sports has not only impacted coaching strategies—ranging from optimizing salary caps to in-game decision-making—but also highlighted discrepancies between perceived and actual risks under different play conditions. Sports data have become a valuable resource in introductory statistics courses and have spurred the development of specialty conferences and publications. For example, the annual MIT Sloan Sports Analytics Conference (SSAC) provides a platform for industry professionals, researchers, students, and enthusiasts to explore the growing influence of analytics in sports.

In basketball, particularly in the National Basketball Association (NBA), statistical methods \citep{sampaio2003statistical,kubatko2007starting,nikolaidis2015building,cervone2016multiresolution} have long been utilized to inform game strategies. Coaches frequently analyze shot data for game preparation and in-game adjustments, with halftime analyses often featuring displays of shot locations and outcomes. While basic statistical techniques offer some insights, the complexity of shot data patterns presents an opportunity for more sophisticated analyses, driving advances in the rapidly evolving field of sports analytics \citep{albert2017handbook}.

An early approach to modeling basketball short chart data through spatial point processes appears in a technical report by \cite{hickson2003spatial}, analyzing shot charts from Michael Jordan during the 2001--2002 NBA season at the Washington Wizards, following his second return from retirement to active play. The dataset includes Jordan's individual shot charts for each game in the season, obtained as $(x,y)$-coordinates from game reports listed on \url{www.espn.com} and \url{www.cbssportsline.com} (data are available as an online supplement). Data include shot outcomes (made/missed), shot location, distances from the basket and side of the court, game played at home or away, and quarter of the game when the shot was taken. In addition to Jordan's shot data, our method development is also motivated by the analysis of shot data of Stephen Curry and LeBron James in the NBA season 2014--2015 from \url{stats.nba.com}. Our goal is to identify the spatial pattern of the missed and made shots and the important factors affecting this pattern. The factors we consider here are whether the game was played at home or away and whether the opponent team went to playoffs (strong) or not (weak) in that season.

Our approach builds on a foundation of previous statistical studies that have modeled basketball shot chart data. \citet{reich2006spatial} introduced a hierarchical Bayesian model with spatially-varying covariates to analyze shot data on a grid of small regions across the half-court. This model employed a multinomial framework to analyze shot counts within grid cells, enabling the assessment of factors influencing shot location choices while accounting for spatial correlation between neighboring grid cells. Building on this, \citet{parker2011modeling} proposed a probit regression model incorporating a conditional autoregressive (CAR) structure to capture spatial variation in regression effects, demonstrating improved accuracy in predicting shot success compared to models that did not account for spatial structure. \citet{goldsberry2012courtvision} introduced CourtVision, a geographic information system-based approach that provided new visual analytic summaries of shot chart data, offering insights into shooting ranges and revealing differences in individual shooting abilities. \citet{lopez2013analisis} applied Kulldorff’s spatial and spatiotemporal scan statistic (SaTScan), typically used in epidemiology, to identify spatial clusters of shots at both team and player levels using shot data from the Los Angeles Lakers from 2007 to 2009. Additionally, \citet{miller2014factorized,yin2020bayesian} analyzed the intensity surface of field goal attempts by professional players, uncovering valuable insights into shooting patterns. Building on intensity estimates, \citet{hu2020bayesiangroup,yin2020analysis} identified latent subgroups among NBA players, shedding light on shooting similarities between different players. \citet{yin2020bayesian} developed a nonparametric Bayesian method to discover heterogeneous shooting patterns of different NBA players based on a nonhomogeneous Poisson process. \citet{jiao2019bayesian} proposed a Bayesian marked point process framework for jointly modeling shot frequency and accuracy, discovering relationships between the two, and \citet{Qi2024AoAS} introduced depth-based testing procedures to test differences between made and missed shots for NBA players.

We build on these developments by modeling the marked point pattern from shot charts as a point process, with results aggregated over small grid cells, focusing on evaluating the local effects of covariates on the probability of NBA players making or missing a shot. Since NBA players generally have a field goal percentage below 50\%, instead of modeling the difference between miss and make intensity functions directly at different covariate levels, we are more interested in identifying areas where a player is more likely to attempt a shot and the associated success risk, referred to as spatial relative risk in Section~\ref{sec:spat_RR}.

Our inferential approach leverages the Log-Gaussian Cox process (LGCP) framework introduced by \cite{moller1998log}. LGCPs provide a flexible structure for making inferences about spatial and spatio-temporal point patterns and have been widely applied across various fields. For instance, \cite{benevs2005case} used an LGCP model with covariates to map disease risk and examine the dependency of risk factors in Central Bohemia, while \cite{diggle2005point} employed a non-stationary LGCP model to analyze spatial-temporal disease surveillance data in Hampshire, UK, and predicted localized deviations from historical disease rates. Similarly, \cite{samartsidis2017bayesian} developed a Bayesian LGCP regression model for meta-analysis of functional neuroimaging data. \cite{kang2011meta} introduced a Bayesian spatial hierarchical model using a marked independent cluster process to model foci as offspring of latent study and population centers, improving inference and capturing inter-study variability. Extending this framework, Kang et al. \cite{kang2014bayesian} proposed a hierarchical Poisson/Gamma random field model for multi-type neuroimaging data, providing a nonparametric approach to handle complex spatial dependencies. In the context of basketball, \cite{miller2014factorized} applied an LGCP model to estimate the intensity function for different players and used non-negative matrix factorization (NMF) techniques to capture distinct shooting patterns across the half-court. Our approach extends this LGCP framework to estimate spatially varying effects of game-specific factors that may influence shot patterns. Building on past methods, we propose a Bayesian hierarchical log Gaussian Cox process model incorporating spatially-varying covariate effects. The hierarchical structure allows our model to borrow information across the half-court during a game and also to borrow information across games to improve estimation precision.  Our method incorporates game-specific covariate effects, leading to local insights on a player's tendencies and abilities. Furthermore, we adopt a kernel convolution approach by using the Karhunen-Loeve expansion, and thus develop an efficient Markov chain Monte Carlo (MCMC) algorithm for posterior inference. By integrating game-specific variables and spatial analytics, our model not only predicts field goal attempts but also identifies the spatially varying influences that affect these attempts. This enhanced modeling approach allows us to provide more nuanced insights into how situational and positional factors impact players' decisions to shoot, thereby contributing valuable perspectives to the strategy and analysis of basketball performance.

The structure of the paper is outlined as follows. Section \ref{sec:data_example} provides an overview of the shooting data for selected players from the National Basketball Association (NBA) regular season. In Section \ref{sec:model}, we introduce the Bayesian Hierarchical Log-Gaussian Cox Process model framework and elaborate on the Bayesian inference procedures, including the Markov Chain Monte Carlo (MCMC) algorithm and the assessment of spatial relative risk. Section \ref{sec:simu} details the comprehensive simulation studies conducted to validate our modeling approach. Section \ref{sec:Analysis} offers a thorough analysis of field goal attempts by three representative NBA players, rigorously analyzing the data to reveal underlying patterns and determinants of shooting decisions. Finally, Section \ref{sec:discussion} concludes the paper with a discussion of the findings and considerations for future research.

\section{The Motivating Data}\label{sec:data_example}

\begin{figure}[h]
    \centering
    \includegraphics[width=\linewidth]{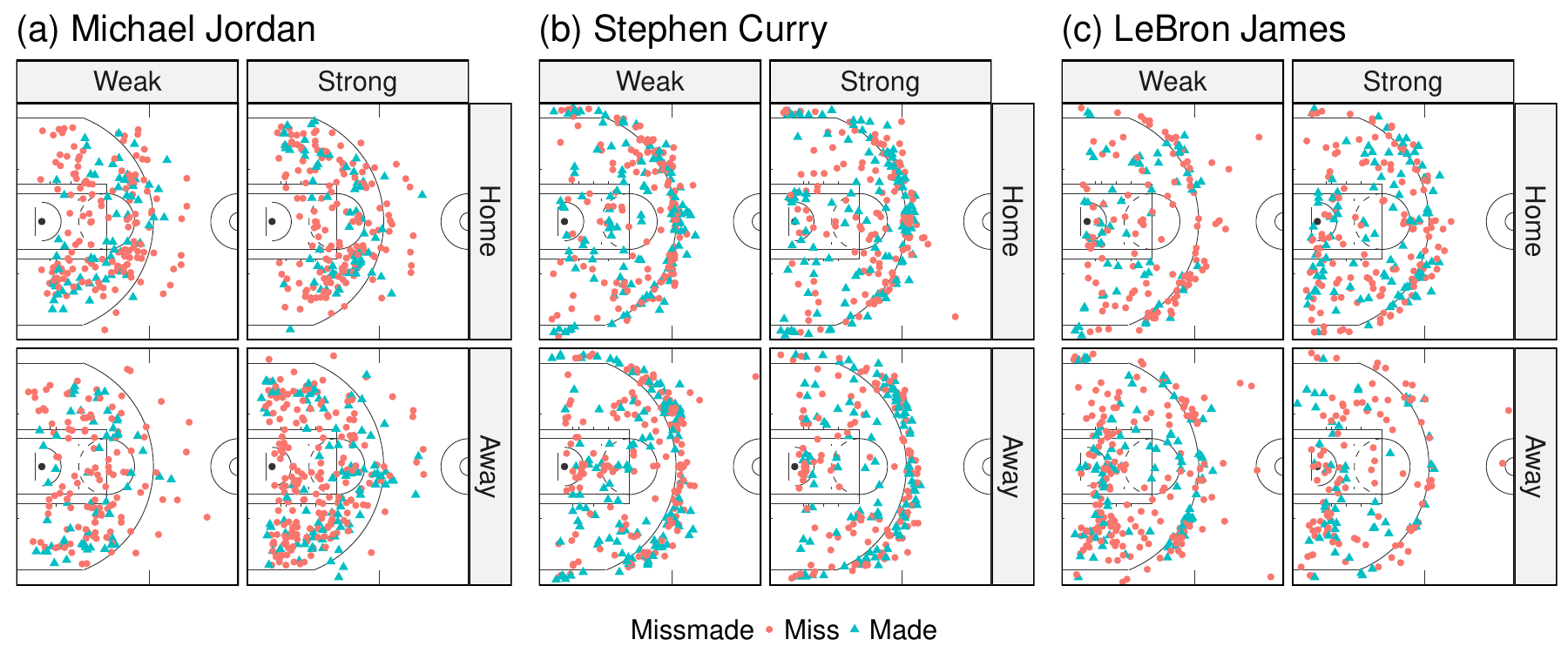}
    \caption{Shot charts of Michael Jordan, Stephen Curry, and LeBron James based on game location (home/away) and opponent team strength (strong/weak).}
    \label{fig:data_illustration}
\end{figure}

The shot charts depicted for Michael Jordan, Stephen Curry, and LeBron James in Figure \ref{fig:data_illustration} showcase distinct shooting patterns and preferences across different game scenarios—specifically when comparing performances in home versus away games and against strong versus weak opponents. Michael Jordan's attempts are predominantly in the mid-range area, aligning with the playing style of his era. In contrast, Stephen Curry’s chart shows a substantial number of attempts from the three-point line, underscoring his impact in shifting the game towards long-range shooting. LeBron James displays a more versatile pattern, with shots evenly distributed between inside the paint and mid-range, indicative of his all-around playing style.

The comparative analysis of these players reveals nuanced strategic adjustments based on the game's context. Both Jordan's and James's shot charts maintain consistent locations regardless of whether the games are home or away, suggesting a steadfast adherence to successful shooting strategies under varied conditions. Curry, however, increases his three-point attempts in away games, possibly to capitalize on high-value shots in less familiar environments. This not only highlights each player's unique strengths and tactical modifications but also illustrates how different eras and styles of play are reflected in individual shooting patterns, providing valuable insights into adapting and optimizing shooting strategies in professional basketball.

Further analysis of the shot charts, especially concerning the impact of playing against weak versus strong opponents, reveals strategic adaptations. Michael Jordan exhibits a higher concentration of mid-range shots against weaker opponents, likely exploiting lower defensive resistance. In contrast, Stephen Curry’s approach against stronger opponents involves an increased frequency of three-point attempts, reflecting a strategic shift to maximize scoring opportunities against tighter defenses that constrict space closer to the basket. This data elucidates how players adjust their tactics based on the strength of the opposition, offering a deeper understanding of game-time decision-making in professional basketball.

Based on the observed shot chart patterns, employing point process model with game-specific covariates for analysis is particularly advantageous. Those models are well-suited to model the intensity function of shots across the basketball court, accounting for the spatial point process nature of shooting data. This method can accurately address the non-uniform distribution of shot attempts and effectiveness across different game situations and player positions. By leveraging the point process model, analysts can quantify the spatial variation in shooting efficiency and explore complex interactions between player actions and game contexts. The adoption of the point process model in this scenario enables a more granular and statistically robust understanding of how external factors like game location and opponent strength impact shooting strategies, thereby providing deeper insights that are crucial for tactical planning and performance enhancement in basketball.

\section{The Model}\label{sec:model}

\subsection{Spatial Point Processes }
We begin with a brief overview of spatial point process analysis via a set of statistical models to capture the characteristics of spatial point patterns. One of the most basic families of point process models is the heterogeneous Poisson point process (PPP),  a stochastic process on a spatial space $\cD$. Suppose $X$ follows a PPP with intensity function $\lambda(x): \cD \longmapsto \mathbb R^+$,  a non-negative function that measures the expectation of the number of observed events in $X$ falling in any subregion in $\cD$. For a PPP, the number of events observed in any $A \subseteq \cD$, denoted $N_A(X)$, follows a Poisson distribution with mean (and variance) $\int_{A} \lambda(x)d x$. Furthermore, if the regions $A$ and $B$ are disjoint, then the random variables $N_A(X)$ and $N_B(X)$ are independent.

An important generalization of the PPP arises when \(\lambda(x)\) itself is treated as a random function. In this case, the process \(X\) is referred to as a \textbf{doubly stochastic Poisson process} or a \textbf{Cox process}. A prominent example of this is the \textbf{log-Gaussian Cox process (LGCP)}, as introduced by \citet{moller1998log}, where \(\log\{\lambda(x)\}\) is modeled as a realization of a Gaussian process (GP). A GP extends the concept of a multivariate normal distribution to infinite-dimensional spaces. It is defined as a stochastic process such that every finite collection of random variables associated with points in the domain follows a multivariate normal distribution \citep{rasmussen2006gaussian}. Formally, a GP is specified by its mean function \(m(x)\) and covariance function \(k(x, x')\), which define the central tendency and correlation structure, respectively, of the random variables over the domain. The flexibility of GPs makes them a powerful modeling tool in statistics and machine learning. They are widely used for applications such as nonparametric regression, where GPs serve as priors over functions; classification, where they provide probabilistic predictions; and spatial modeling, where they capture spatially correlated random effects. Within LGCPs, the Gaussian process allows for modeling the latent spatial variation in \(\lambda(x)\), enabling the analysis of more complex point patterns, including those exhibiting clustering or heterogeneity. Overall, the combination of Poisson processes with Gaussian processes in LGCPs provides a rich framework for modeling spatial point patterns, balancing interpretability and flexibility while accommodating diverse real-world phenomena.

\subsection{\bf A  Hierarchical Log-Gaussian Cox Process Model}\label{prior}

Let $\mbR^d$ represent a $d$-dimensional Euclidean space ($d\geq 1$). Suppose we consider $m$ games. Let $i (i = 1,\ldots,m)$ index game. Let $j (j=0,1)$ index the type of shots, where $j=0$ represents the missed shot and $j=1$ represents the made shot. For game $i$, let $\bfs_{i,j} = \{\bfs_{i,j,t}\}_{t=1}^{t_{i,j}}$ denote a collection of the type $j$ shots in half court region $\cR := \{(x, y)\} \subset \mbR^2$, that is, $\bfs_{i,j} \subset \cR$. In addition, we collect covariate $\bfz_i = (z_{i,1},\ldots, z_{i,p})^{\rT}\in \mbR^p$ for game $i$.     Let $\pp(A,\lambda)$ denote a Poisson point process on region $A$ with intensity $\lambda$.   We consider observed basketball shot patterns as realizations of a hierarchical LGCP model as follows.
\begin{eqnarray}
(\bfs_{i,j} \mid \lambda_{j}, \bfz_i )&\sim& \pp\{\cR, \lambda_{j}(\bfs; \bfz_i) \}, \label{eq:a}
\end{eqnarray}
where $\lambda_{j}(\bfs; \bfz_i)$ represents the intensity function for type $j$ shots at location $\bfs$ for the game with covariate $\bfz_i$. We assume
\begin{eqnarray}
\log\{\lambda_{j}(\bfs;\bfz_i)\} = \alpha_0(\bfs) +  \bfz_i^{\rT}\bfbeta_j(\bfs),
\end{eqnarray}
where spatially-varying intercept $\alpha_0(\bfs)$ represents the baseline log intensity of shot patterns across all games and two different types at location $\bfs$. The spatially-varying coefficient $\bfbeta_j(\bfs) = \{\beta_{j,1}(\bfs),\ldots, \beta_{j,p}(\bfs)\}^{\rT}$ represents the spatial-effects of the covariates $\bfz_i$ at location $\bfs$ for type $j$ shots.  More specifically, after adjusting the covariate effects,  ``$\exp\{\alpha_0(\bfs)\}$" represents the population level local expected number of shots across all games and two different types. According to \cite{moller2003statistical}, the probability density of shot patterns $\bfS = \{\{\bfs_{i,j}\}_{i=1}^m\}_{j=0}^1$ with parameter $\bftheta$ and covariates $\bfz = \{\bfz_i\}_{i=1}^m$, is given by
\begin{eqnarray}
\pi\left(\bfS \mid \lambda_0(\cdot,\cdot),\lambda_1(\cdot,\cdot)\right)= \prod_{i,j} \left\{\exp\left\{|\cR| - \int_{\cR} \lambda_{j}(\bfs; \bfz_i)d \bfs \right\} \prod_{\bfs\in \bfs_{i,j}} \lambda_{j}(\bfs; \bfz_i)\right\}.
\end{eqnarray}

We assign Gaussian process priors for all the spatially-varying coefficients in the model. Let $\gp(\mu, k)$ denote a Gaussian process with mean  $\mu$ and covariance kernel $k$. We assume $\alpha_0(\bfs)$ and $\beta_{j,k}(\bfs)$ are {\it a priori} mutually  independent of each other and 
\begin{eqnarray*}
\alpha_0(\bfs) \sim \gp(0,\sigma^2_0\kappa), \qquad \beta_{j,k}(\bfs) \sim \gp(0,\sigma^2_{j,k} \kappa),
\end{eqnarray*}
for $i = 1,\ldots, m$, $j = 0,1$ and $k = 1,\ldots, p$. To be simple,  all the covariate effects are assumed to share the same prior correlation kernel $\kappa$ but with different prior variances. 

By \emph{Mercer's theorem} \citep{rasmussen2006gaussian}, the covariance kernel $\kappa(\bfs,\bfs')$ can be decomposed as $\kappa(\bfs,\bfs')= \sum_{l=1}^{\infty}\xi_l \phi_l(\bfs)\phi_l(\bfs')$, where $\{\xi_l\}_{l=1}^{\infty}$ denote the eigenvalues and $\{\phi_l(\bfs)\}_{l=1}^{\infty}$ the eigenfunctions. These satisfy $\xi_l\geq \xi_{l+1}$ for $l\geq 1$, $\int \phi_l(\bfs)\phi_{l'}(\bfs) d s =0, \forall l \neq l'$ , $\int \kappa(\bfs,\bfs') \phi_l(\bfs) d s = \xi_l \phi_{l}(\bfs)$ and $\int \phi_l(\bfs) d s =1$ for any $l$. This further implies that the spatial processes $\alpha_0(\bfs)$ and $\beta_{j,k}(\bfs)$ can be represented as:
\begin{eqnarray*}
\alpha_0(\bfs) = \sum_{l=1}^\infty \theta^0_l\phi_l(\bfs),\mbox{with}\  \theta^0_l \sim \mN(0,\sigma^2_0\xi_l), \mbox{ and }  \beta_{j,k}(\bfs) = \sum_{l=1}^\infty \theta^{\beta}_{j,k,l}\phi_l(\bfs),\mbox{with}\   \theta^\beta_{j,k,l} \sim \mN(0,\sigma^2_{jk}\xi_l).
\end{eqnarray*}
where $\xi_l$ is determined by kernel function. Here we set $\sigma^2_{jk} = \sigma^2_{\beta}$ for all $j\in\{0,1\}$ and $k\in\{1,2.\ldots,p\}$, and consider the kernel $\kappa(\bfs,\bfs') = \exp\{-a (\|\bfs\|^2+\|\bfs'\|^2)-b\|\bfs-\bfs'\|^2\}$, which is a commonly used kernel for Gaussian Processes modelling. Of course, other kernel functions such as Mat\'ern kernel could also be used. See \citep{rasmussen2006gaussian} for other kernel functions used in Gaussian Processes modeling.
Thus, we can approximate $\alpha_0(\bfs)$ and $\beta_{j,k}(\bfs)$ using a finite number of eigenfunctions. That is
\begin{eqnarray*}
\alpha_0(\bfs) \approx \sum_{l=1}^L \theta^0_l\phi_l(\bfs)= \bm{\phi}^{\rT}(\bfs)\bm{\theta}^{0}, \quad \mbox{and}\quad \beta_{j,k}(\bfs) \approx \sum_{l=1}^L \theta^{\beta}_{j,k,l}\phi_l(\bfs)=
 \bm{\phi}^{\rT}(\bfs)\bm{\theta}^{\beta}_{j,k},
\end{eqnarray*}
where $\bm{\phi}(\bfs)=[\phi_1(\bfs),\ldots,\phi_L(\bfs)]^{\rT},\bm{\theta}^{0}=(\theta^0_1,\ldots,\theta^0_L)^{\rT},
\bm{\theta}^{\beta}_{j,k}=({\theta}^{\beta}_{j,k,1},\ldots,{\theta}^{\beta}_{j,k,L})^{\rT}$, $j=0,1$, and $k=1,\ldots, p$. The number of component $L$ is usually chosen such that $\sum_{l=1}^{L}\xi_l/\sum_{l=1}^{\infty}\xi_l > \alpha$ for a given $\alpha\in (0,1)$.

Then we have
\begin{eqnarray*}
\log[\lambda_{j}(\bfs; \bfz_i)] = \bfX_{i,j}^{\rT}(\bfs) \bftheta = \bm{\phi}(\bfs)^{\rT}\bm{\theta}^{0}+(\bfz_i^{\rT}\otimes\bfphi(\bfs)^{\rT}){\bftheta}^{\beta}_{j},
\end{eqnarray*}
where  $\bfX_{i,j}^{\rT}(\bfs)=(1,c_j\otimes z^{\rT}_i)\otimes \bfphi(\bfs)^{\rT}$ and ``$\otimes$" denotes the Kronecker product.  $c_j$ is the $(j+1)$th row of $I_2, j=0,1$, $\bftheta = ({\bftheta}^{0\rT},{\bftheta}^{\beta\rT}_{0},{\bftheta}^{\beta\rT}_{1})^{\rT}$, with $\bm{\theta}^{0}=(\theta^0_1,\ldots,\theta^0_L)^{\rT}$ and $\bm{\theta}^{\beta}_j=(\bm{\theta}^{\beta\rT}_{j,1},\ldots,\bm{\theta}^{\beta\rT}_{j,p})^{\rT}$, for $j=0,1$. $\bftheta \sim \mN(\bfzero,\bfSigma)$ with
$\bfSigma = \diag\{\sigma^2_0,\sigma^2_{\beta}I_{2p}\}\otimes \Xi$ and $\Xi=\diag \{\xi_1,\ldots,\xi_L\}$.

\begin{figure}[h]
    \centering
    \includegraphics[width=\linewidth]{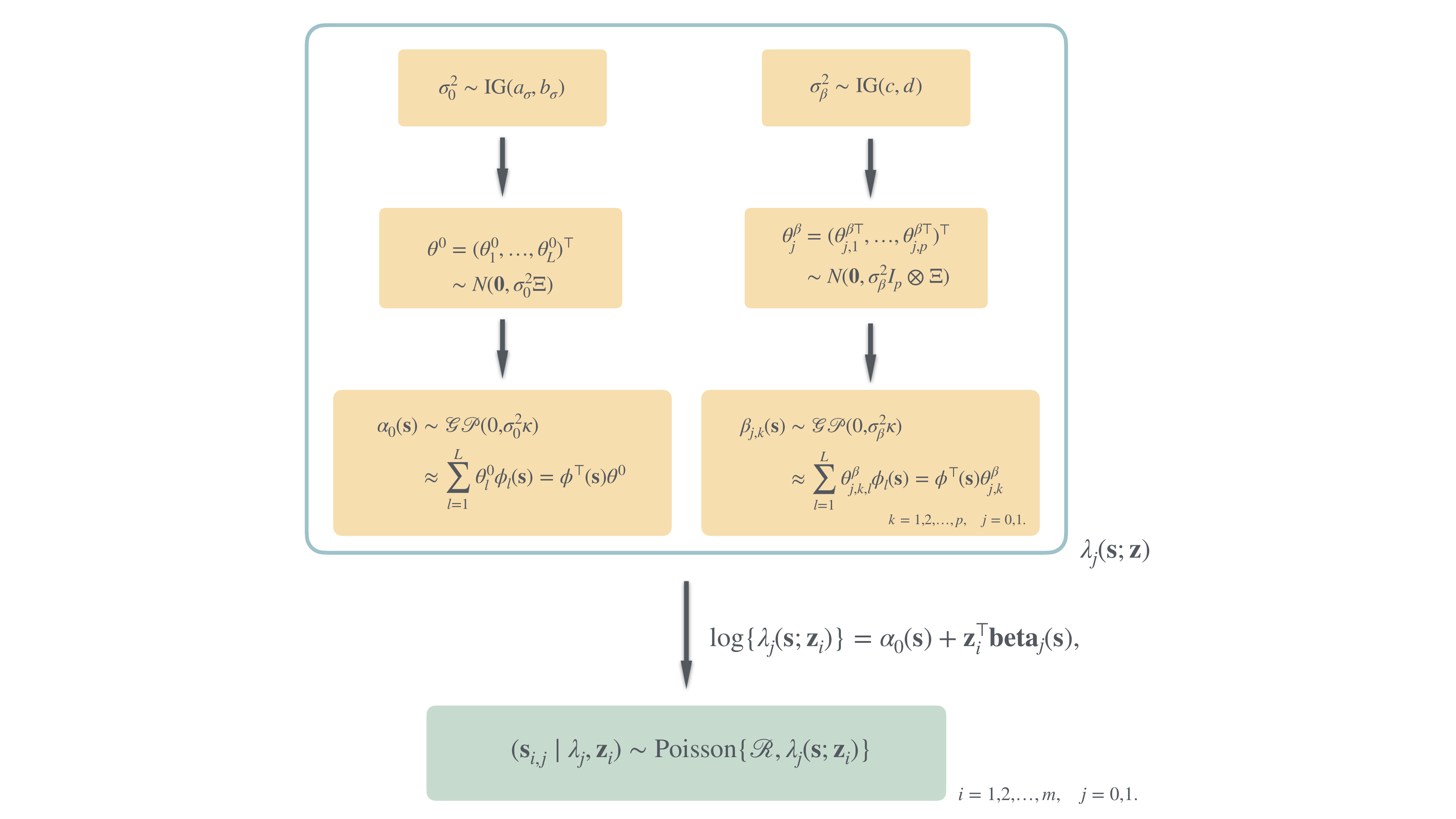}
    \caption{Model diagram of the proposed Jointly Spatial-Varying Log-Gaussian Cox Process Model, with priors highlighted in orange rectangles and likelihoods in the green rectangle.}
    \label{fig:model_diagram}
\end{figure}

Finally, we assign an inverse-Gamma priors to the hyper-parameters $\sigma_0^2$ and $\sigma_{\beta}^2$,  i.e.,  $\sigma_0^2\sim \mIG(a_\sigma,b_\sigma)$ and  $\sigma_{\beta}^2\sim IG(c,d)$ with positive parameters $a_\sigma,b_\sigma, c$, and $d$. The proposed model is summarized in Figure~\ref{fig:model_diagram}.

\subsection{Posterior Inference}\label{sec:post_inference}

The posterior distribution of $\bftheta$ given the observed locations $\bfS$ is
\begin{eqnarray*}
\lefteqn{\pi(\bftheta^0, \{\bftheta^{\beta}_j\}_{j=0}^1\mid \bfS) \propto \pi(\bfS \mid \bftheta) \pi(\bftheta) }\\
 &&\propto\exp\left[- \sum_{i,j}\int_{\cR} \exp\left\{\bfX_{i,j}^{\rT}(\bfs)\bftheta \right\}\md \bfs+\sum_{i=1}^{m} \sum_{j=0}^{1} \sum_{t=1}^{t_{i,j}} \bfX_{i,j}^{\rT}(\bfs_{i,j,t})\bftheta-\frac{1}{2}\bftheta^{\rT}\bfSigma^{-1}\bftheta \right].
 \end{eqnarray*}
The full conditional distribution of $\bftheta^0$ given the other parameters and data $\bfS$ is
\begin{eqnarray*}
\lefteqn{\pi(\bftheta^0 \mid \bullet, \sigma_0^2,\bfS) }\\
&& \propto \exp\left[- \sum_{i,j}\int_{\cR} \exp\{\bm{\phi}^{\rT}\bm{\theta}^{0}+(\bfz_i^{\rT}\otimes\bfphi^{\rT}){\bftheta}^{\beta}_{j}\}\md \bfs+\sum_{i=1}^{m} \sum_{j=0}^{1} \sum_{t=1}^{t_{i,j}}\bfphi^{\rT}(\bfs_{i,j,t})\bftheta^0-\frac{1}{2}\bftheta^{0\rT}(\sigma_0^2\Xi)^{-1}\bftheta^0 \right].
\end{eqnarray*}
The full conditional of $\sigma_0^2$ given the other parameters and data $\bfS$ is
$$\left [ \sigma_0^2 \mid \bftheta^0,\bullet\right ] \sim \mIG\left[a_\sigma+\frac{L}{2},b_\sigma+\frac{1}{2}\bftheta^{0\rT}\Xi^{-1}\bftheta^0\right].$$
We update $\bftheta^0$ via the Metropolis-Hastings algorithm. However, the acceptance ratio requires the integral of the intensity function over the entire region which has no closed form. We therefore adopt an approximation to calculate the integral as follows. For equally-spaced grid cells $\{\bfs^{(h)} = (\theta_{h},\delta_{h})\}_{h=1}^H$ on region $\cR$ in the Cartesian coordinate system,  we approximate the integral by
\begin{eqnarray*}
\int_{\cA} \exp\left\{\bfX_{i,j}^{\rT}(\bfs)\bftheta \right\} \md \bfs \approx \sum_{h=1}^{H} \exp\left\{\bfX_{i,j}^{\rT}(\bfs^{(h)})\bftheta \right\} \|\bfs^{(h)}\|.
\end{eqnarray*}
where $\|\bfs^{(h)}\|$ is the grid cell area.

To speed up the convergence of the Markov chain, we adopt the Metropolis-Adjusted Langevin Algorithm \citep[MALA]{roberts1998optimal}. Let $\phi(\bfx\mid \bfmu,\bfSigma)$ denote the density function of a multivariate normal distribution with mean $\bfmu$ and $\bfSigma$.

Specifically, our algorithm for updating $\bftheta^0$ and $\sigma^2_0$ is as follows:
\begin{itemize}
\item Input: shot pattern $\bfS = \{\{\bfs_{i,j}\}_{i=1}^m\}_{j=0}^1$.
\item Set $\bftheta^{0(1)} =\diag(\bfI_L,\bfzero)\left(\sum_{j=0}^1\sum_{i=1}^m \widetilde{\bfX}_{i,j}^{\rT}\widetilde{\bfX}_{i,j}\right)^{-1}\sum_{j=0}^1\sum_{i=1}^m \widetilde{\bfX}_{i,j}^{\rT}\log(m_{i,j}/s\md x\md y)$,
where $\widetilde{\bfX}_{i,j}=\sum_{t=1}^{t_{i,j}} \bfX_{i,j}^{\rT}(\bfs_{i,j,t}) $ where $s\md x\md y$ denotes the area of a single grid cell and $m_{i,j}$ is the number of type $j$ points in game $i$.
\item Draw $\sigma_0^{2(1)} \sim \mIG(a_\sigma,b_\sigma)$.
\item Repeat the following steps for $v=1,2,\ldots,N$,
\begin{itemize}
\item Draw $\bfzeta^0\sim \mN[\mu_0(\bftheta^{0(v)}),\tau_0^2 I_L]$ and $u\sim \mU(0,1)$. Set
 \begin{eqnarray*}
\lefteqn{ \mu_0(\bftheta^{0(v)})=\bftheta^{0(v)}+\frac{\tau_0^2}{2}\left[-\sum_{i,j,h} \exp\{{\bm\phi}^{\rT}(\bfs^{(h)})\bm{\theta}^{0(v)}+(\bfz_i^{\rT}\otimes\bfphi^{\rT}){\bftheta}^{\beta}_{j}\}\bm{\phi}(\bfs^{(h)})\right.}\\
 && \qquad\qquad \qquad\qquad\qquad \left.+\sum_{i=1}^{m} \sum_{j=0}^{1} \sum_{t=1}^{t_{i,j}}\bfphi(\bfs_{s,j,t})-(\sigma_0^{2(v)}\Xi)^{-1}\bm{\theta}^{0(v)}\right].
  \end{eqnarray*}
\item If $u<  \frac{\pi(\bfzeta^0 \mid \bullet, \sigma_0^{2(v)},\bfs)q_0(\bftheta^{0(v)}\mid \bfzeta^0)}{\pi(\bftheta^{0(v)} \mid \bullet, \sigma_0^{2(v)},\bfs)q_0(\bfzeta^0 \mid \bftheta^{0(v)})}$,
then set $\bftheta^{0(v+1)}=\bfzeta^0$, else set $\bftheta^{0(v+1)}=\bftheta^{0(v)}$, where $q_0(\bftheta^{0(v)}\mid \bfzeta^0)=\phi\{\bftheta^{0(v)}\mid \mu_0(\bfzeta^0),\tau_0^2 I_L\}$.
\item Draw $\sigma_0^{2(v+1)} \sim \mIG[a_\sigma+\frac{L}{2},b_\sigma+\frac{1}{2}\bftheta^{0(v+1)\rT}\Xi^{-1}\bftheta^{0(v+1)}]$.
\end{itemize}
\item Output: $\{\bftheta^{0(1)},\bftheta^{0(2)},\ldots,\bftheta^{0(N)}\}; \{\sigma_0^{2(1)},\sigma_0^{2(2)},\cdots,\sigma_0^{2(N)}\}$.
\end{itemize}
We develop MALA algorithms and Gibbs samplers for updating  $\{\bftheta^{\beta}_j\}_{j=0}^1$ and $\sigma_{\beta}^2$ respectively. Please refer to the Supplementary Section~S.1 for details.

\subsection{Spatial relative risk}\label{sec:spat_RR}
Typically, NBA players have a field goal percentage less than 50\%. Thus, to better show the different intensity patterns under different covariates values, we consider the spatial relative risk at different locations. Following \cite{hickson2003spatial}, we adopt the relative risk notation from epidemiology to define the relative risk of a player making a shot at different covariates values across different locations. In epidemiology, relative risk is typically defined as the ratio of the probability of diseases in the exposed group to the probability of diseases in the control group. \cite{kelsall1995non} considered the ratio of two intensity estimates as the spatial measure of relative risk. To define the spatial relative risk for basketball games, we consider the following spatially-varying event $\mathcal{T}(\bfs) = \{\mbox{a shot taken at }\bfs\}$, $\mathcal{T}_1=\{\mbox{a made shot taken at }\bfs\}$ and $\mathcal{T}_0 = \{\mbox{a miss shot taken at }\bfs\}$.  Clearly, we have $\mathcal{T}(\bfs) = \mathcal{T}_1(\bfs) \cup \mathcal{T}_0(\bfs)$ and $\mathcal{T}_0(\bfs)\cap \mathcal{T}_1(\bfs) = \emptyset$ for any $\bfs \in \cR$. 
Then we define the spatial relative risk at location $\bfs$ between two covariate values $\bfa$ and $\bfb$, denoted $\mathrm{RR}(\bfs; \bfa,\bfb)$,  as the ratio of the probability of a made shot at $\bfs$ given the covariate taking the two different values correspondingly. 

\begin{equation}\label{eq:RR}
\mathrm{RR}(\bfs; \bfa,\bfb) = \frac{\mbox{Pr}\{\cT_1(\bfs) \mid \bfz = \bfa\}}{\mbox{Pr}\{\cT_1(\bfs) \mid \bfz = \bfb\}}.
\end{equation}
We can compare $\mathrm{RR}(\bfs; \bfa,\bfb)$ with one to determine whether the player has a higher, equal or lower ``risk" to take a made shot at location $\bfs$ between the covariate $\bfz = \bfa$ and $\bfz = \bfb$.  

We can represent $\mathrm{RR}(\bfs; \bfa,\bfb)$ using the intensity functions in the proposed model. 
Given covariate $\bfz$, the probability of a made shot taken at $\bfs$, i.e., $\mbox{Pr}\{\cT_1(\bfs) \mid \bfz \}$ can be decomposed as the probability of a shot taken at $\bfs$, i.e., $\mathrm{Pr}\{\cT(\bfs) \mid \bfz\}$, multiplied by the probability of the shot being a made shot at $\bfs$, i.e. $\mathrm{Pr}\{\cT_1(\bfs) \mid \cT(\bfs), \bfz\}$,   In short, 
\begin{equation}\label{eq:T_1}
\mbox{Pr}\{\cT_1(\bfs) \mid \bfz \} = \mathrm{Pr}\{\cT(\bfs) \mid \bfz\} \mathrm{Pr}\{\cT_1(\bfs) \mid \cT(\bfs), \bfz\}. 
\end{equation}
According to our model, the intensity function $\lambda_j(\bfs;\bfz)$ measures the expectation of the number of type $j$ shots taken at location $\bfs$. Thus, $\mathrm{Pr}\{\cT(\bfs) \mid \bfz\}$ is proportional to the expectation of number of missed and made shots taken at location $\bfs$ given covariate $\bfz$, where the normalizing constant is equal to  the expectation of number of missed and made shots taken across half court region $\cR$ given covariate $\bfz$. 
\begin{align}\label{eq:T}
\mathrm{Pr}\{\cT(\bfs) \mid \bfz\}  = \frac{\lambda_1(\bfs;\bfz) +\lambda_0(\bfs;\bfz)}{\int_{\cR}[\lambda_1(\bfs';\bfz)+\lambda_0(\bfs';\bfz) ]d \bfs'}. 
\end{align}
After adjusting covariate $\bfz$, the conditional probability of a shot being the made shot at location $\bfs$ can be represented as the expected number of made shots taken at location $\bfs$  divided by the expected number of missed and made shots taken at location $\bfs$. That is, 
\begin{align}\label{eq:T_1_T}
\mathrm{Pr}\{\cT_1(\bfs) \mid \cT(\bfs), \bfz\} &= \frac{\lambda_1(\bfs;\bfz)}{\lambda_1(\bfs;\bfz)+\lambda_0(\bfs;\bfz)}.
\end{align}
By combining \eqref{eq:T_1}--\eqref{eq:T_1_T}, \eqref{eq:RR} becomes
$$\mathrm{RR}(\bfs; \bfa,\bfb)  = \frac{\lambda_1(\bfs;\bfa)}{\lambda_1(\bfs;\bfb)}\times \frac{\int_{\cR}[\lambda_1(\bfs';\bfb)+\lambda_0(\bfs';\bfb)] d \bfs'}{\int_{\cR}[\lambda_1(\bfs';\bfa)+\lambda_0(\bfs';\bfa) ]d \bfs'}.$$

\section{Simulations}\label{sec:simu}

In this section, we investigate the performance of the proposed method using simulated data. The data are generated according to the model described in Section~\ref{sec:model}. We consider two scenarios: one where the model parameter $\bm{\theta}$ is randomly generated (denoted by ``Synthetic Intensity"), and another where it is chosen to mimic the spatial shot pattern of Stephen Curry (denoted by ``Curry Intensity").

Specifically, we simulate data for $m = 200$ games, with $50$ games for each combination of game location (home or away) and opponent team level (strong or weak). Under the first scenario, ``Synthetic Intensity", the eigenfunctions $\phi_l$ are specified by choosing $a = b = 1$ and $L = 3$. The elements in the model parameter $\bm{\theta} \in \mathbb{R}^{7L}$ are randomly sampled from a uniform distribution on $[-1,1]$. Under the second scenario, ``Curry Intensity", the eigenfunctions are specified as those adopted in Section~\ref{sec:Analysis}, and the true $\bm{\theta}$ is set to the posterior mean estimate fitted from the shot data of Stephen Curry based on the last $5,000$ MCMC iterations.

In each scenario, we fit the proposed Jointly Spatial-Varying Log-Gaussian Cox Process Model (denoted by ``JSVLGCP") alongside several competitive methods to demonstrate the superiority of the proposed model. Specifically, we consider the standard Log-Gaussian Cox Process model (denoted by ``LGCP")  implemented with the R package \textit{inlabru}, and the nonhomogeneous Poisson process model (denoted by ``IPP"), where the intensity functions are modeled with cubic B-spline functions and implemented with the R package \textit{spatstat}. We also implement kernel density estimation (denoted by ``KDE") and Bayesian Additive Regression Trees (denoted by ``BART") using the R packages \textit{spatstat} and \textit{BART}, respectively. Performance is evaluated using the root mean square error (RMSE), defined as
\begin{equation}
    \text{RMSE} = \sqrt{\frac{1}{N} \sum_{i=1}^m \sum_{j=0}^1 \sum_{t=0}^{t_{i,j}} \Big[\hat{\lambda}_j(\bfs_{i,j,t}, \bm{z}_j) - \lambda_j(\bfs_{i,j,t}, \bm{z}_j)\Big]^2},
\end{equation}
where $\hat{\lambda}_j(\bfs_{i,j,t}, \bm{z}_j)$ denotes the predicted intensity value at the observed shot location, and $N$ denotes the total number of observations. For Bayesian methods, point estimation is performed using the posterior mean estimates.

\begin{figure}
    \centering
    \includegraphics[width=0.9\linewidth]{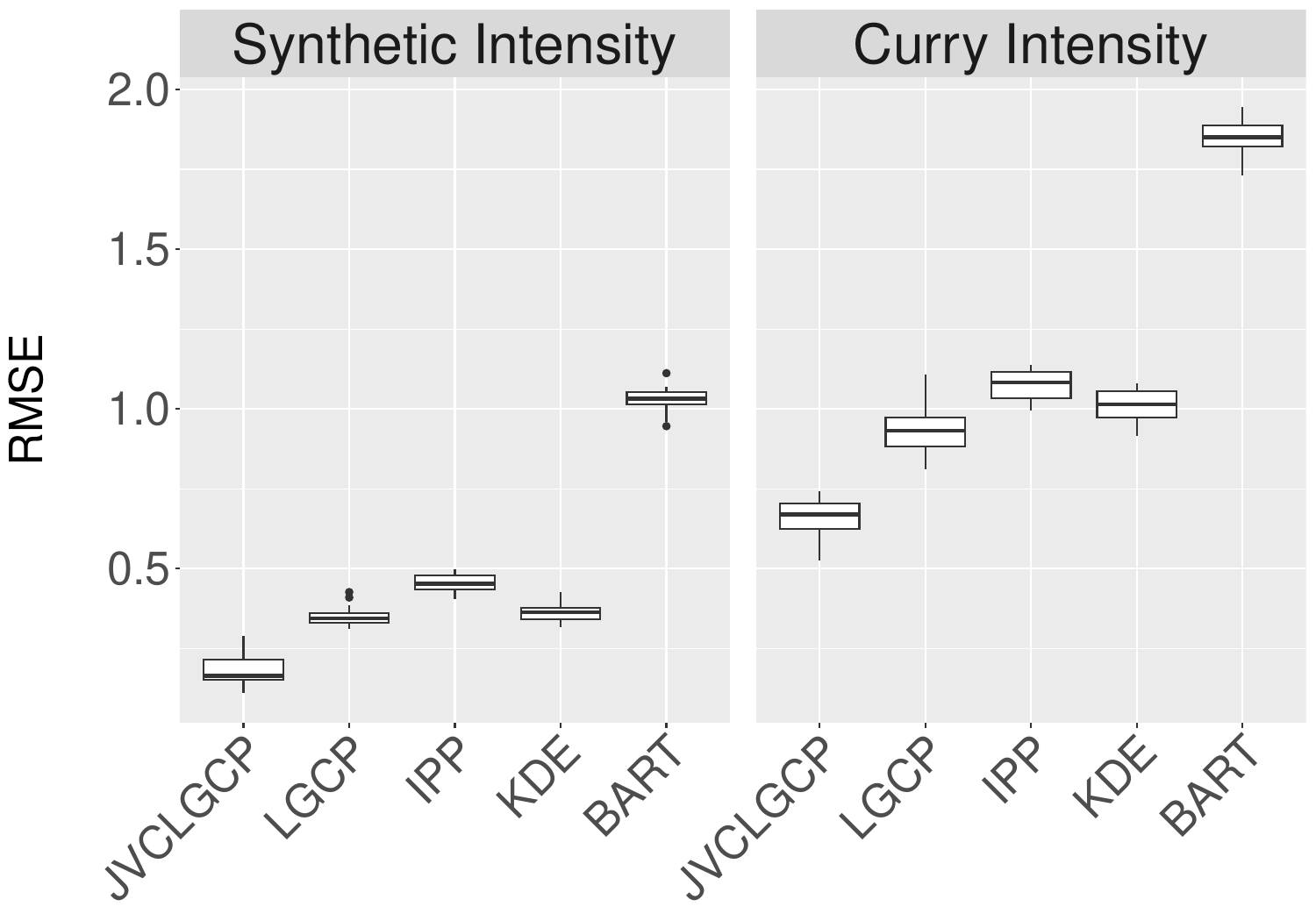}
    \caption{Model comparison based on RMSE.}
    \label{fig:simu:boxplot}
\end{figure}

\begin{figure}
    \centering
    \includegraphics[width=1\linewidth]{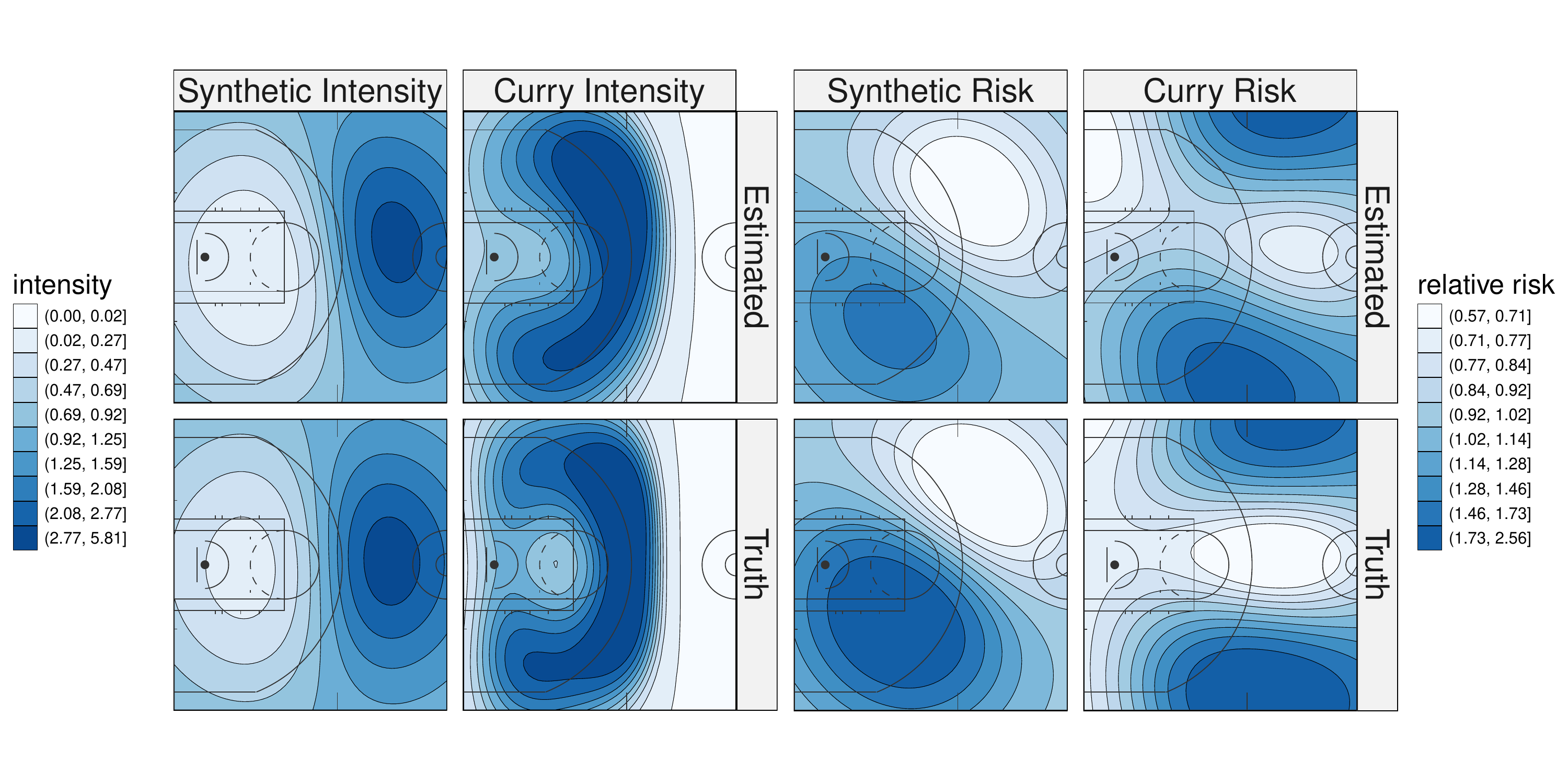}
    \caption{JVCLGCP estimation of intensity (left) and relative risk (right).}
    \label{fig:simu:summary}
\end{figure}

The estimation results are reported in Figure~\ref{fig:simu:boxplot}, where we present the RMSE results in side-by-side boxplots based on 20 replicate datasets for each scenario. The proposed JVCLGCP consistently outperforms all four competitive methods, yielding accurate estimates for the underlying intensity values. To further demonstrate the estimated intensity patterns of JVCLGCP, we compare the estimated spatially varying intensity and relative risk (denoted by ``Estimated") with their respective true specifications (denoted by ``Truth") in Figure~\ref{fig:simu:summary}. In particular, we report the made intensity functions for home team against strong opponents, and the relative risk between home vs away for strong opponents. The results indicate that our approach provides accurate estimates of both the intensity functions and the spatially varying relative risk.

\section{Analysis of Shot Charts data}\label{sec:Analysis}
We apply the model to analyze shooting data from NBA players during specific seasons. This includes examining shots taken by Stephen Curry and LeBron James in the 2014-2015 season, and Michael Jordan's shots from the 2001-2002 season. We exclude shots that are more than 28 feet from the basket, as these are often taken in the final moments of quarters or games, and also very short-range attempts like lay-ups and slam dunks that are less than 1 foot from the basket. This allows us to focus on typical shooting patterns within games and to observe more nuanced differences in the likelihood of making a shot based on shot location.

For Stephen Curry and LeBron James, the shot data were retrieved from the website stats.nba.com. Each record in the dataset specifies the action type associated with a shot. We excluded shots categorized as layups and slam dunks from our analysis. After these adjustments, we retained data on 1,066 shots across 80 games for Stephen Curry and 856 shots across 69 games for LeBron James. For Michael Jordan, we removed 2 shots that were taken from beyond 28 feet and 285 shots that were less than 1 foot from the basket, focusing on more standard field goal attempts. This left us with a dataset comprising 908 shots in 54 games for our analysis of Jordan's performance. In our analysis, we included covariates such as the location of the game (home or away) \citep{mateus2021home} and the strength of the opponent \citep{south2024basketball}, categorized based on whether the opposing team qualified for the playoffs (strong) or not (weak) in that particular season. These factors are considered to potentially influence shooting performance and are included to adjust for their effects in our statistical models.

We make inferences on the shot intensity functions on a set of equal spaced grid cells within the half-court region. In the MALA algorithm, initial values are set as follows: $\tau_0^{2}=0.03/[{L(p+1)}]^{1/3}$, $\tau_{\beta}^{2}=0.03/[{L(p+1)}^{1/3}]$, $\sigma_0^2 \sim \mIG(5,5)$ and $\sigma_\beta^2 \sim \mIG(5,5)$. The length $L$ of the eigenvalue vector for the exponential kernel is determined by the values of $a$, $b$ and  the percentage of the intensity variance explained after the truncation, denoted $\alpha$. We fix $a=0.25$ and set $b =1.5$, which is determined by maximizing the marginal likelihood. We set $L=15$ such that the recovery rate of the KL expansion $\sum_{l=0}^{L}\xi_l/\sum_{l=0}^{\infty}\xi_l > 0.8$. The MCMC algorithm was run for 15,000 iterations with first $10,000$ as burn-in.

\begin{figure}[htbp]
  \centering
  \includegraphics[height=0.9\textheight]{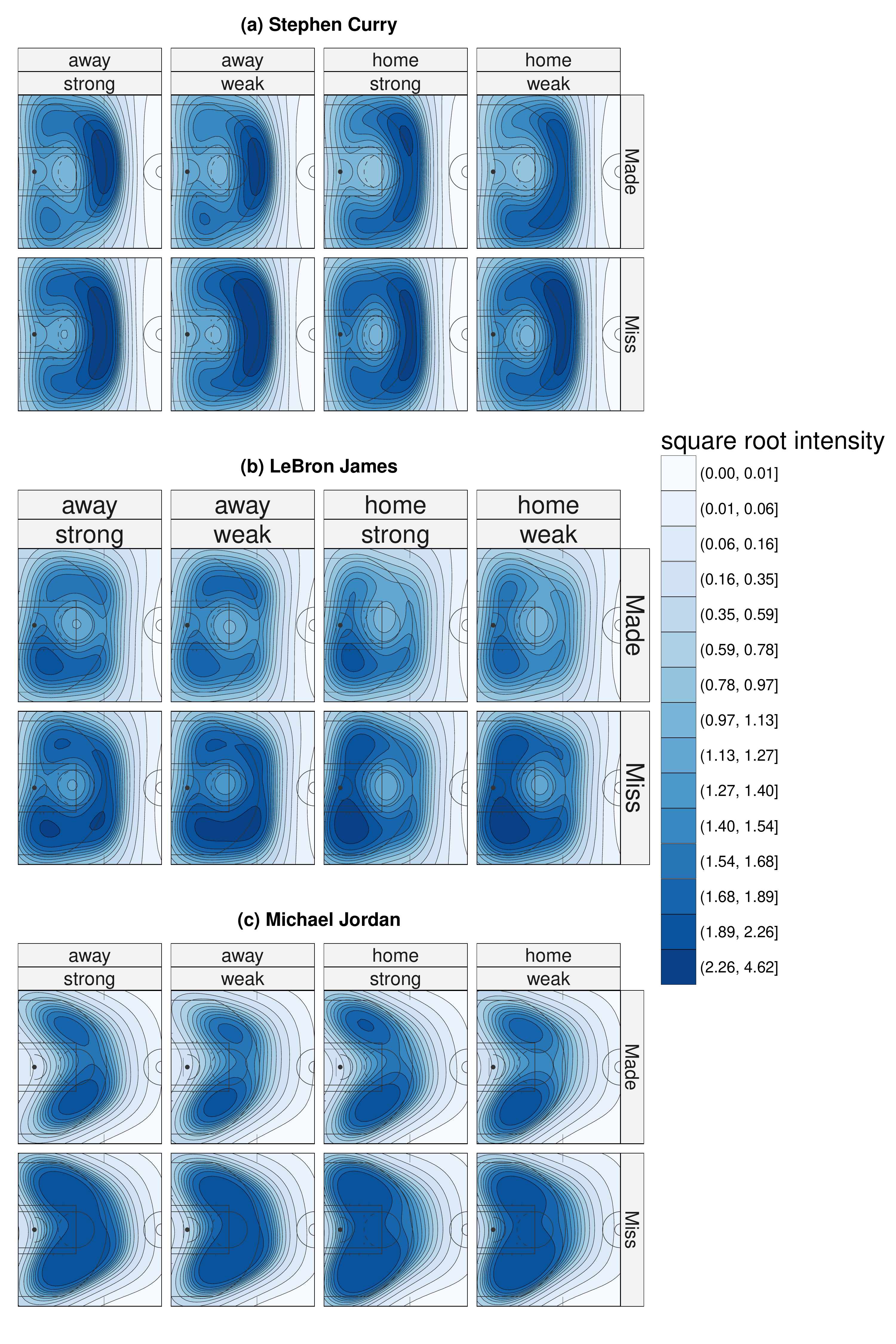}
\caption{Map of the square root of the estimated intensity function for different players under different covariate values.}
\label{intensity} 
\end{figure}

Figure \ref{intensity} illustrates the estimated intensity function for missed and made shots, adjusted for different covariate values on a half basketball court. We have transformed the intensity function values by taking their square roots to enhance the visibility of differences on the same scale. The figure distinctly reveals the varied shooting patterns among the three players. Curry, renowned for his three-point shooting skills, shows distinct patterns in Figure \ref{intensity} (a). Both the intensity of his missed and made shots peak near the three-point line directly facing the basket, highlighting his proficiency and preference for long-range shots. In contrast, Figure \ref{intensity} (b) displays LeBron James' tendencies, where he is most effective under the basket. His shooting patterns vary significantly depending on the game situation and opposing team strength, emphasizing his adaptability and strength in close-range scoring. Figure \ref{intensity} (c) showcases Michael Jordan's balanced shooting strategy. Jordan's shot-making ability is demonstrated to be symmetrical around the basket, but with a noticeably higher intensity on the left side under all examined conditions. This pattern underscores Jordan's strategic use of space and his effectiveness in utilizing both sides of the court, although with a slight preference for the left. Overall, these visualizations provide a comprehensive comparison of each player's strategic approach to shot-making, influenced by their unique strengths and the dynamics of each game.

\begin{figure}[t]
  \centering
  \includegraphics[width=\textwidth]{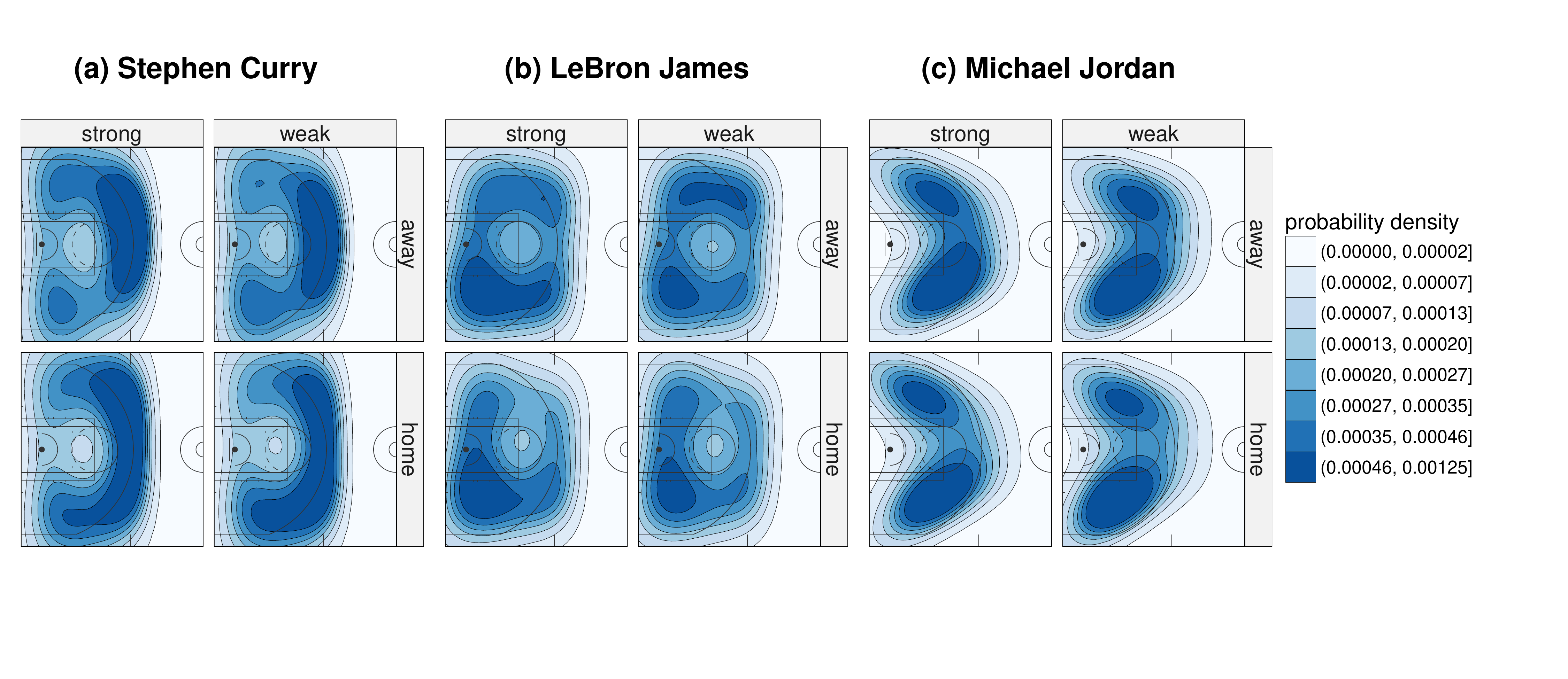}
\vspace{-1cm}
\caption{Map of the estimated probability density function for different players under different covariate values. The estimated probability for each subfigure sums to 1.}
\label{risk} 
\end{figure}

Figure \ref{risk} presents the estimated probability density for a successful shot at locations under varying covariate conditions for different players, providing insights into each player's preferred shooting areas and their effectiveness in those regions. For Stephen Curry (Figure \ref{risk}(a)), high probability regions predominantly align along the arcs of the three-point line, underscoring his reputation as an exceptional long-range shooter. This pattern emphasizes Curry's ability to consistently convert shots from beyond the arc, making him a formidable opponent in perimeter shooting. LeBron James' shot probability (Figure \ref{risk}(b)) exhibits consistent patterns across different game scenarios, with the highest probabilities concentrated near the basket. Interestingly, these high-probability regions expand when facing stronger opponents compared to weaker ones. This expansion suggests that James is particularly effective at driving to the basket and scoring in high-pressure situations against top-tier defenses, likely using his physicality and skill to navigate through tighter defenses. Michael Jordan's chart (Figure \ref{risk}(c)) illustrates high probability densities that are almost symmetrical around the basket, stretching from the three-point line to the free-throw line. The probability density is notably higher on the left side of the court. This symmetry and the left-side dominance might reflect Jordan's strategic play and shooting preferences, possibly indicating favored spots for taking critical shots. Together, these visualizations provide a detailed comparative analysis of each player's shooting efficacy and strategic preferences on the court, highlighting how different conditions and opponents influence their game-time decisions and outcomes.

\begin{figure}[htbp]
  \centering
  \includegraphics[width = 0.9\textwidth]{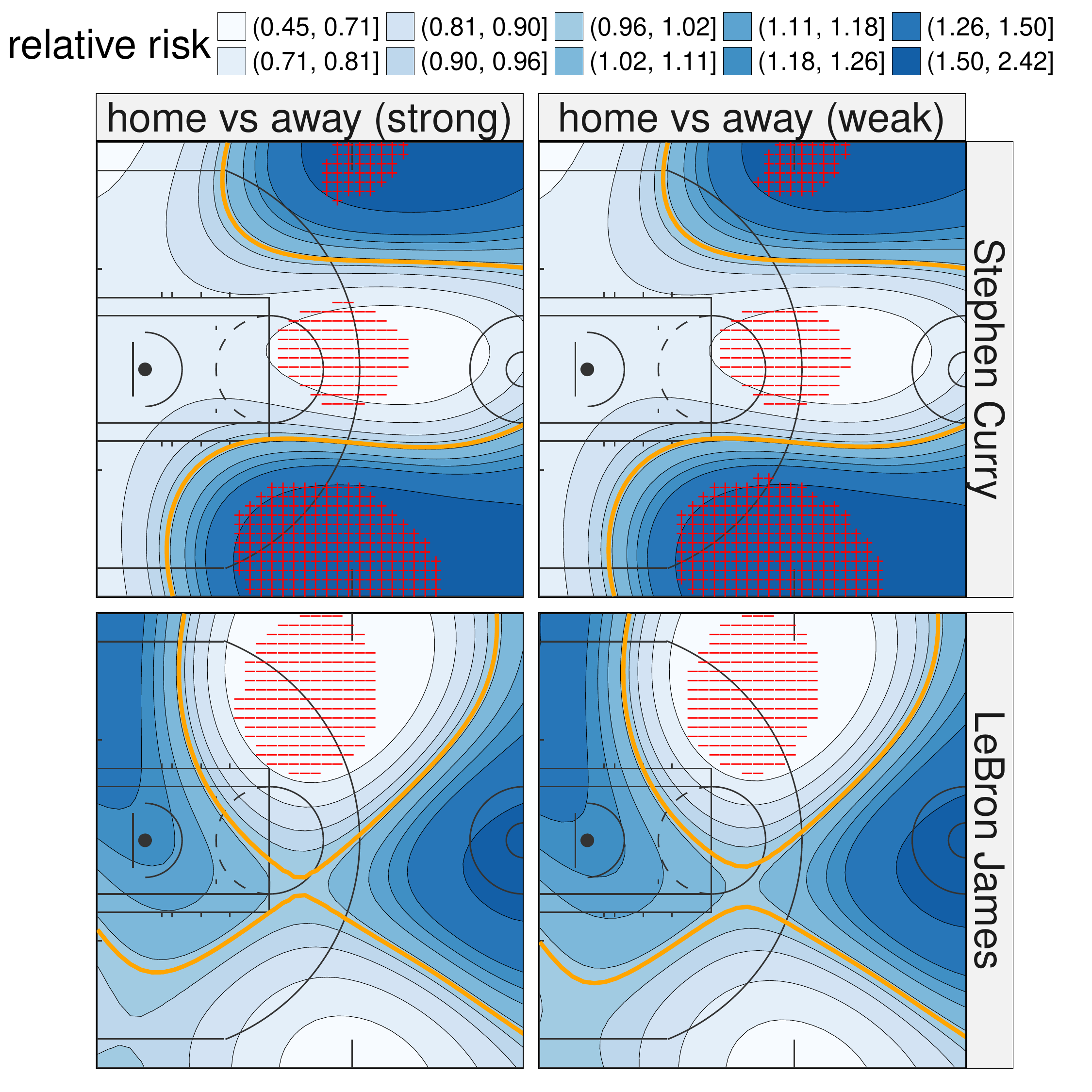}
\caption{Map of the estimated home vs away relative risk under different covariate values for Stephen Curry and LeBron James. '+' means the 90\% CI of the risk between made intensity and miss intensity is significantly larger than 1. '-' means the 90\% CI of the risk between made intensity and miss intensity is significantly smaller than 1. Orange line is the contour line for relative risk as 1.}
\label{curry_james} 
\end{figure}

Figure \ref{curry_james} delves into the nuances of playing environments by comparing the estimated relative risks of shooting success for Stephen Curry and LeBron James in home versus away games against both strong and weak opponents. This analysis not only highlights different performance patterns based on location but also incorporates the 90\% credible intervals (CI) for these spatial relative risks, marked by the symbols '+' and '-' across the visualizations, to underscore significant differences. For Stephen Curry, the analysis reveals distinct advantages at specific court locations when playing at home. Notably, the top two panels of Figure \ref{curry_james} indicate higher relative risks near both the left and right corners of the three-point arc, suggesting that Curry's accuracy from these spots is significantly better in home settings. Conversely, there is a noticeable area between the free throw line and the three-point line directly facing the basket where Curry's performance dips in home games compared to away games, highlighting a unique aspect of his shooting pattern that might be influenced by varying court dynamics or psychological factors at his home arena. LeBron James exhibits a different pattern, with increased relative risks occurring near the top of the three-point line and along the right baseline. These findings suggest that James leverages different areas of the court effectively depending on the home field. Intriguingly, the bottom panels of Figure \ref{curry_james} reveal that James performs significantly better in an area around the middle of the right three-point line when playing away games. This unusual pattern suggests that unlike many players, James might not benefit as much from the typical home field advantage, possibly due to his adaptability and consistent performance under varying crowd pressures and court surroundings. These insights into Curry and James's performances under different conditions offer a deeper understanding of how environmental factors such as crowd support and familiar settings influence player effectiveness. Such analyses not only enrich the basketball narrative by discussing strategic deployments based on venue but also help coaches and analysts fine-tune player utilization strategies in crucial game situations.
\begin{figure}[htbp]
  \centering
  \includegraphics[width = 0.8\textwidth]{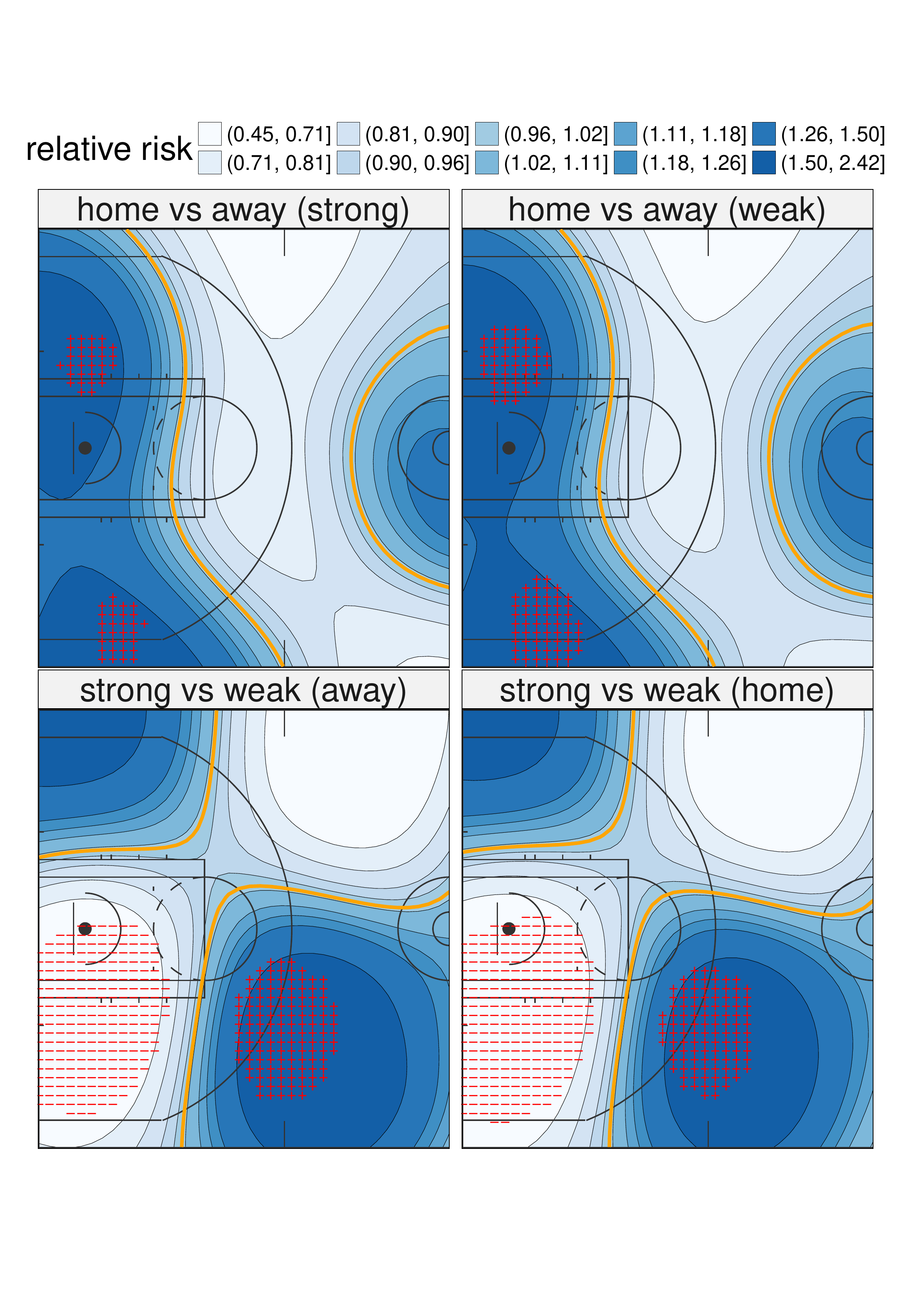}
  \vspace{-2cm}
\caption{Map of the estimated relative under different covariate values for Michael Jordan. '+' means the 90\% CI of the risk between made intensity and miss intensity is significantly larger than 1. '-' means the 90\% CI of the risk between made intensity and miss intensity is significantly smaller than 1. Orange line is the contour line for relative risk as 1.}
\label{jordan_risk} 
\end{figure}

Figure \ref{jordan_risk} explores the estimated relative risks under various covariate values for Michael Jordan, showcasing distinct patterns that diverge significantly from those observed for Stephen Curry and LeBron James. For Jordan, the analysis indicates a higher relative risk of successful shooting outcomes in different court locations depending on game settings and opponent strength. In the top two panels of Figure \ref{jordan_risk}, we see that Jordan's performance is significantly better in specific areas when playing home games as opposed to away games, while accounting for the strength of the opposing team. One notable area of improved performance is around the left bottom of the three-point line, suggesting Jordan's adeptness in utilizing this region effectively in familiar settings. Another key area is directly beside the basket on the right bottom of the court, where his ability to score increases markedly during home games. These regions demonstrate Jordan's tactical use of his home court advantage, emphasizing areas where he feels most comfortable and likely experiences less defensive pressure. Furthermore, the bottom two panels of Figure \ref{jordan_risk} reveal how Jordan's performance fluctuates significantly depending on whether he faces strong or weak opponents. Against strong opponents, Jordan's shooting from between the free throw line and the three-point line on the left middle of the court becomes more effective. This adaptation likely stems from the increased defensive focus around the basket by stronger teams, forcing Jordan to exploit mid-range areas that might be less heavily defended. Conversely, when playing against weaker opponents, his performance peaks near the basket on the left bottom side of the court, exploiting the lesser defensive capabilities in these areas.

These findings suggest that Jordan adapted his playing style to maximize scoring opportunities based on the defensive characteristics of his opponents. This adaptability not only highlights his strategic acumen but also offers valuable insights for opposing coaches in devising defensive strategies tailored to mitigate his impact. The contrast in the playing styles and strategic adaptations between Jordan, Curry, and James also reflects the evolution of basketball strategies over the years, indicating shifts in how players and teams leverage home field advantages and respond to the defensive setups of their adversaries.

\begin{figure}
    \centering
    \includegraphics[width=0.9\linewidth]{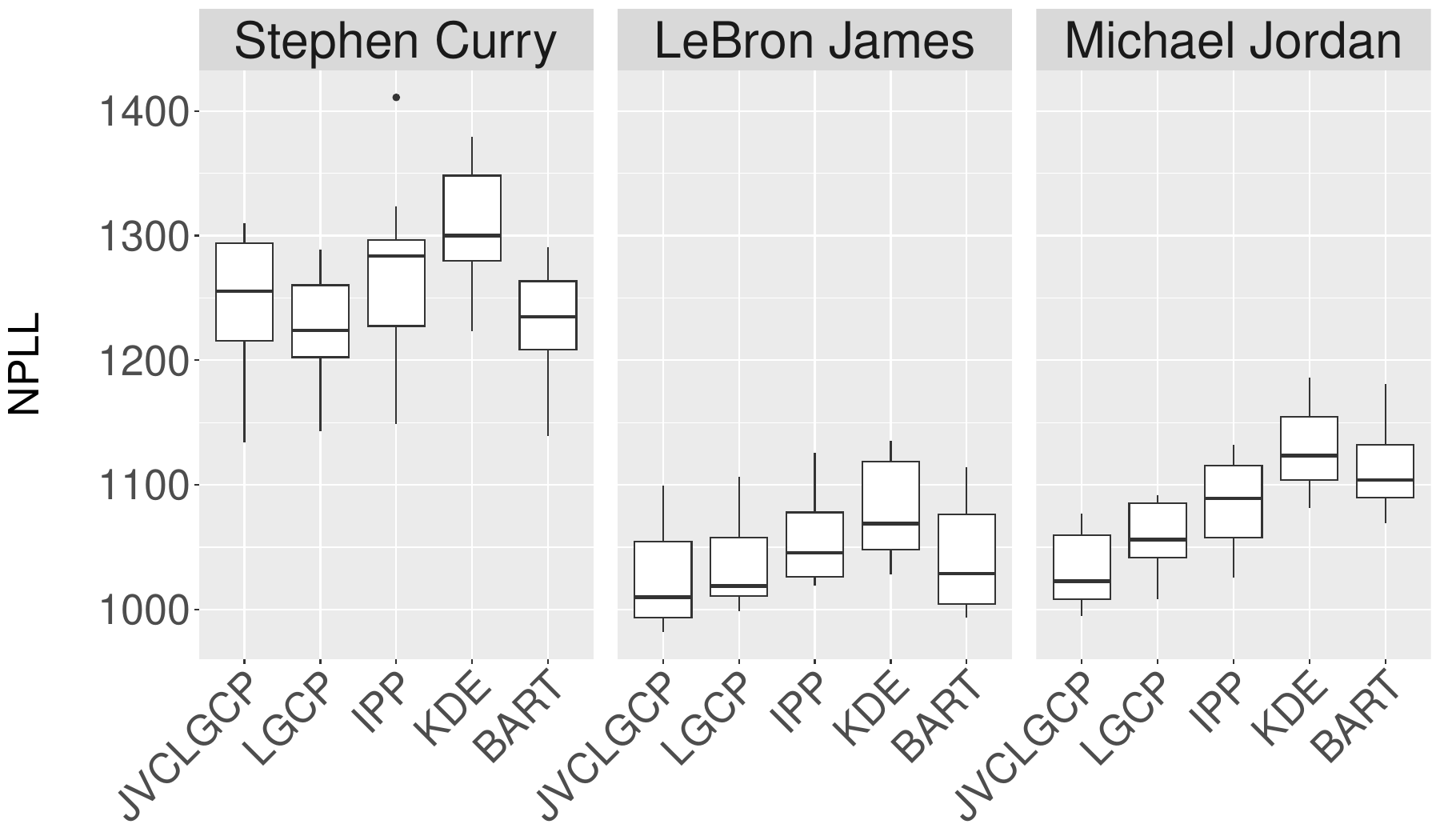}
    \caption{Model comparison based on NPLL.}
    \label{fig:realdata:boxplot}
\end{figure}

Finally, we conduct a model comparison on the real shot data of the three players between the proposed JVCLGCP with all the competitive methods introduced in Section~\ref{sec:simu}. To evaluate model prediction performance, we introduce the Negative Predictive Log Likelihood (NPLL) based on $p$-thinning. We first partition the spatial domain into 400 rectangular regions, $S_{l} \subseteq \mathbb{R}$, for $l = 1, 2, \ldots, 400$. Let $p \in (0, 1)$ denote the retention probability. The $p$-thinning procedure involves selecting each shot data point for the training data independently with probability $p$, while the remaining data points form the testing data. We fit the models on the training data and then calculate the NPLL on the testing data, defined as
\begin{equation}
    \text{NPLL} = -\sum_{i=1}^m \sum_{j=0}^1 \sum_{\bfz} \sum_{l=1}^{400}
    \log\left[p\left(n^*_{S_l,i,j,\bfz} \Big\vert \frac{1-p}{p} \hat{\lambda}_{S_l,i,j,\bfz} \right)\right],
\end{equation}
where $\hat{\lambda}_{S_l,i,j,\bfz}$ is the estimated intensity of region $S_{l}$ given covariate $\bfz$, type $j$, and game $i$, and $n^*_{S_l,i,j,\bfz}$ is the corresponding number of observed shots falling into region $S_{l}$ in the test data. We fix $p = 0.8$, as per \citet{geng2019bayesian}, and note that the model class with a smaller NPLL value fits the data better from a prediction perspective. The above procedure is independently performed $10$ times, and the NPLL results are presented in Figure~\ref{fig:realdata:boxplot}. Compared with the competitive models, we find that JVCLGCP achieves the lowest NPLLs on the shot data of LeBron James and Michael Jordan, and yields comparable results on the shot data of Stephen Curry.

\section{Discussion}\label{sec:discussion}
In the sections above, we outlined and executed a Log-Gaussian Cox Processes (LGCP) model with spatially varying coefficients, aimed at dissecting basketball shot chart data. This methodology establishes a solid basis for elucidating spatial patterns in shot outcomes—differentiating made from missed shots—while accounting for covariate influences that can differ by location. The inherent flexibility of this model is derived from the application of Gaussian Processes (GPs) which effectively capture spatial dependencies and model spatially varying coefficients, thus embracing the intricate spatial configurations prevalent in basketball shot data. Additionally, we have advanced the use of a fast Metropolis-adjusted Langevin algorithm (MALA) to facilitate efficient posterior simulation and inference. This innovation enables the model's scalable application across extensive datasets, with simulation studies corroborating the accuracy and dependability of our estimation process, solidifying the practicality of our approach.

The LGCP framework, enhanced with spatially varying coefficients, possesses extensive potential applications within the realm of sports analytics. Its capacity to intricately model spatial distributions, integrate dynamic covariate effects, and adjust to temporal variations renders it an invaluable tool for analyzing varied sports-related datasets. In soccer, for example, the model could be used to estimate goal probabilities based on shot locations, taking into account the defensive formations at the time of each shot. In tennis, the model could analyze serve success rates across different zones of the court, potentially revealing patterns that could guide players' serve placement strategies.

Moreover, by integrating spatially varying defensive covariates, the model can provide insights into how defensive alignments influence shot outcomes, thus aiding teams in refining their defensive tactics. This analytical framework can be further extended to incorporate game-context variables such as clutch moments or high-pressure situations, offering a deeper understanding of how athletes' performances fluctuate under varying game conditions. These capabilities demonstrate the model's robust applicability and potential to revolutionize strategic planning and performance analysis across sports disciplines..

The proposed LGCP framework sets the stage for numerous innovative developments in sports analytics. Future research could enrich this framework by enhancing spatio-temporal modeling to more accurately capture the evolving dynamics of player and team behaviors during matches and over successive seasons. Such enhancements would deepen the understanding of temporal patterns and strategic shifts in gameplay. Furthermore, the framework could be augmented with advanced machine learning methods, such as deep learning algorithms, to adeptly handle the complex, non-linear interactions among spatial configurations and various covariates. Incorporating data from wearable sensors that monitor player movements and physiological responses could significantly enrich the analysis of spatio-temporal interactions and their impact on player performance and fatigue levels. To address data scarcity, the implementation of regularization techniques or hierarchical models could enhance the robustness of the analyses, especially in areas with sparse data or infrequent occurrences. Future studies might also explore clustering and classification methods specifically tailored to contexts of shot selection. By classifying players into distinct groups based on their shooting styles and preferences, and clustering shot types by outcome efficiency and situational factors, teams could develop more targeted strategies for player training and game planning. These methods would not only refine the LGCP framework but also transform raw data into strategic insights that could revolutionize player recruitment and tactical planning in professional basketball. Additionally, integrating the "Expected Points Above Average" (EPAA) metric \citep{williams2024expected} as the marks of the proposed point process model will provide a more granular analysis of player effectiveness. This metric allows for comparisons across different eras and styles of play, adjusting for changes in game pace and defensive strategies over time, offering a nuanced, data-driven insight into the true impact of a player's offensive contributions relative to league averages.

\bibliography{ref}

\begin{thebibliography}{30}
\newcommand{\enquote}[1]{``#1''}
\expandafter\ifx\csname natexlab\endcsname\relax\def\natexlab#1{#1}\fi

\bibitem[{Albert et~al.(2017)Albert, Glickman, Swartz, and Koning}]{albert2017handbook}
Albert, J., Glickman, M.~E., Swartz, T.~B., and Koning, R.~H. (2017), \textit{Handbook of statistical methods and analyses in sports}, Crc Press.

\bibitem[{Bene{\v{s}} et~al.(2005)Bene{\v{s}}, Bodl{\'a}k, M{\o}ller, and Waagepetersen}]{benevs2005case}
Bene{\v{s}}, V., Bodl{\'a}k, K., M{\o}ller, J., and Waagepetersen, R. (2005), \enquote{A case study on point process modelling in disease mapping,} \textit{Image Analysis and Sterology}, 24, 159--168.

\bibitem[{Cervone et~al.(2016)Cervone, D’Amour, Bornn, and Goldsberry}]{cervone2016multiresolution}
Cervone, D., D’Amour, A., Bornn, L., and Goldsberry, K. (2016), \enquote{A multiresolution stochastic process model for predicting basketball possession outcomes,} \textit{Journal of the American Statistical Association}, 111, 585--599.

\bibitem[{Diggle et~al.(2005)Diggle, Rowlingson, and Su}]{diggle2005point}
Diggle, P., Rowlingson, B., and Su, T.-l. (2005), \enquote{Point process methodology for on-line spatio-temporal disease surveillance,} \textit{Environmetrics}, 16, 423--434.

\bibitem[{Geng et~al.(2021)Geng, Shi, and Hu}]{geng2019bayesian}
Geng, J., Shi, W., and Hu, G. (2021), \enquote{Bayesian nonparametric nonhomogeneous Poisson process with applications to USGS earthquake data,} \textit{Spatial Statistics}, 41, 100495.

\bibitem[{Goldsberry(2012)}]{goldsberry2012courtvision}
Goldsberry, K. (2012), \enquote{Courtvision: New visual and spatial analytics for the nba,} in \textit{2012 MIT Sloan Sports Analytics Conference}.

\bibitem[{Hickson et~al.(2003)Hickson, Waller, et~al.}]{hickson2003spatial}
Hickson, D., Waller, L.~A., et~al. (2003), \enquote{Spatial analyses of basketball shot charts: An application to Michael Jordans 2001-2002 NBA season,} in \textit{Proceedings of the Hawaii International Conference on Statistics and Related Fields, Available online at: http://www. hicstatistics. org/2003StatsProceedings/DeMarc\% 20Hickson. pdf (accessed 9 June 2006)}.

\bibitem[{Hu et~al.(2020)Hu, Yang, and Xue}]{hu2020bayesiangroup}
Hu, G., Yang, H.-C., and Xue, Y. (2020), \enquote{Bayesian group learning for shot selection of professional basketball players,} \textit{Stat}, e324.

\bibitem[{Jiao et~al.(2021)Jiao, Hu, and Yan}]{jiao2019bayesian}
Jiao, J., Hu, G., and Yan, J. (2021), \enquote{A Bayesian marked spatial point processes model for basketball shot chart,} \textit{Journal of Quantitative Analysis in Sports}, 17, 77--90.

\bibitem[{Kang et~al.(2011)Kang, Johnson, Nichols, and Wager}]{kang2011meta}
Kang, J., Johnson, T.~D., Nichols, T.~E., and Wager, T.~D. (2011), \enquote{Meta Analysis of Functional Neuroimaging Data via Bayesian Spatial Point Processes,} \textit{Journal of the American Statistical Association}, 106, 124--134.

\bibitem[{Kang et~al.(2014)Kang, Nichols, Wager, and Johnson}]{kang2014bayesian}
Kang, J., Nichols, T.~E., Wager, T.~D., and Johnson, T.~D. (2014), \enquote{A Bayesian Hierarchical Spatial Point Process Model for Multi-Type Neuroimaging Meta-Analysis,} \textit{The Annals of Applied Statistics}, 8, 1800--1824.

\bibitem[{Kelsall and Diggle(1995)}]{kelsall1995non}
Kelsall, J.~E. and Diggle, P.~J. (1995), \enquote{Non-parametric estimation of spatial variation in relative risk,} \textit{Statistics in Medicine}, 14, 2335--2342.

\bibitem[{Kubatko et~al.(2007)Kubatko, Oliver, Pelton, and Rosenbaum}]{kubatko2007starting}
Kubatko, J., Oliver, D., Pelton, K., and Rosenbaum, D.~T. (2007), \enquote{A starting point for analyzing basketball statistics,} \textit{Journal of quantitative analysis in sports}, 3.

\bibitem[{L{\'o}pez et~al.(2013)L{\'o}pez, Mart{\'\i}nez, and Ruiz}]{lopez2013analisis}
L{\'o}pez, F., Mart{\'\i}nez, J., and Ruiz, M. (2013), \enquote{An{\'a}lisis espacial de lanzamientos en baloncesto; el caso de LA Lakers,} \textit{Revista Internacional de Medicina y Ciencias de la Actividad F{\'\i}sica y del Deporte/International Journal of Medicine and Science of Physical Activity and Sport}, 13, 585--613.

\bibitem[{Mateus et~al.(2021)Mateus, Gon{\c{c}}alves, and Sampaio}]{mateus2021home}
Mateus, N., Gon{\c{c}}alves, B., and Sampaio, J. (2021), \enquote{Home advantage in basketball,} in \textit{Home Advantage in Sport}, Routledge, pp. 211--219.

\bibitem[{Miller et~al.(2014)Miller, Bornn, Adams, and Goldsberry}]{miller2014factorized}
Miller, A., Bornn, L., Adams, R., and Goldsberry, K. (2014), \enquote{Factorized Point Process Intensities: A Spatial Analysis of Professional Basketball.} in \textit{ICML}, pp. 235--243.

\bibitem[{M{\o}ller et~al.(1998)M{\o}ller, Syversveen, and Waagepetersen}]{moller1998log}
M{\o}ller, J., Syversveen, A.~R., and Waagepetersen, R.~P. (1998), \enquote{Log gaussian cox processes,} \textit{Scandinavian journal of statistics}, 25, 451--482.

\bibitem[{Moller and Waagepetersen(2003)}]{moller2003statistical}
Moller, J. and Waagepetersen, R.~P. (2003), \textit{Statistical inference and simulation for spatial point processes}, CRC Press.

\bibitem[{Nikolaidis(2015)}]{nikolaidis2015building}
Nikolaidis, Y. (2015), \enquote{Building a basketball game strategy through statistical analysis of data,} \textit{Annals of Operations Research}, 227, 137--159.

\bibitem[{Parker(2011)}]{parker2011modeling}
Parker, R.~J. (2011), \enquote{Modeling College Basketball Shot Location Data,} \textit{Technical Report}.

\bibitem[{Qi et~al.(2024)Qi, Hu, and Wu}]{Qi2024AoAS}
Qi, K., Hu, G., and Wu, W. (2024), \enquote{{Are made and missed different? An analysis of field goal attempts of professional basketball players via depth based testing procedure},} \textit{The Annals of Applied Statistics}, 18, 2615 -- 2634.

\bibitem[{Rasmussen and Williams(2006)}]{rasmussen2006gaussian}
Rasmussen, C. and Williams, C. (2006), \textit{Gaussian Processes for Machine Learning}, Adaptative computation and machine learning series, University Press Group Limited.

\bibitem[{Reich et~al.(2006)Reich, Hodges, Carlin, and Reich}]{reich2006spatial}
Reich, B.~J., Hodges, J.~S., Carlin, B.~P., and Reich, A.~M. (2006), \enquote{A spatial analysis of basketball shot chart data,} \textit{The American Statistician}, 60, 3--12.

\bibitem[{Roberts and Rosenthal(1998)}]{roberts1998optimal}
Roberts, G.~O. and Rosenthal, J.~S. (1998), \enquote{Optimal scaling of discrete approximations to Langevin diffusions,} \textit{Journal of the Royal Statistical Society: Series B (Statistical Methodology)}, 60, 255--268.

\bibitem[{Samartsidis et~al.(2017)Samartsidis, Wager, Barrett, Atzil, Eickhoff, Nichols, and Johnson}]{samartsidis2017bayesian}
Samartsidis, P., Wager, T.~D., Barrett, L.~F., Atzil, S., Eickhoff, S.~B., Nichols, T.~E., and Johnson, T.~D. (2017), \enquote{Bayesian log-Gaussian Cox process regression with applications to fMRI meta-analysis,} \textit{arXiv preprint arXiv:1701.02643}.

\bibitem[{Sampaio and Janeira(2003)}]{sampaio2003statistical}
Sampaio, J. and Janeira, M. (2003), \enquote{Statistical analyses of basketball team performance: understanding teams’ wins and losses according to a different index of ball possessions,} \textit{International Journal of Performance Analysis in Sport}, 3, 40--49.

\bibitem[{South(2024)}]{south2024basketball}
South, C. (2024), \enquote{A basketball paradox: exploring NBA team defensive efficiency in a positionless game,} \textit{Journal of Quantitative Analysis in Sports}.

\bibitem[{Williams et~al.(2024)Williams, Schliep, Fosdick, and Elmore}]{williams2024expected}
Williams, B., Schliep, E.~M., Fosdick, B., and Elmore, R. (2024), \enquote{Expected Points Above Average: A Novel NBA Player Metric Based on Bayesian Hierarchical Modeling,} \textit{arXiv preprint arXiv:2405.10453}.

\bibitem[{Yin et~al.(2023)Yin, Hu, and Shen}]{yin2020analysis}
Yin, F., Hu, G., and Shen, W. (2023), \enquote{Analysis of professional basketball field goal attempts via a Bayesian matrix clustering approach,} \textit{Journal of Computational and Graphical Statistics}, 32, 49--60.

\bibitem[{Yin et~al.(2022)Yin, Jiao, Yan, and Hu}]{yin2020bayesian}
Yin, F., Jiao, J., Yan, J., and Hu, G. (2022), \enquote{Bayesian nonparametric learning for point processes with spatial homogeneity: A spatial analysis of NBA shot locations,} in \textit{International Conference on Machine Learning}, PMLR, pp. 25523--25551.

\end{thebibliography}
\end{document}